\documentclass[12pt]{amsart}
\usepackage[margin = 1in]{geometry}
\usepackage[shortlabels]{enumitem}
\usepackage{graphicx}
\usepackage{subfig}
\usepackage{hyperref}
\usepackage{commath}

\newtheorem{theorem}{Theorem}
\newtheorem{lemma}{Lemma}
\theoremstyle{definition}
\newtheorem{definition}{Definition}
\newtheorem{asmp}{Assumption}

\begin{document}
\title[A New Stock Market Valuation Measure]{A New Stock Market Valuation Measure with Application to Retriement Planning}

\author{Andrey Sarantsev}

\address{University of Nevada, Reno, Department of Mathematics and Statistics}
\email{asarantsev@unr.edu}

\begin{abstract} 
We generalize the classic Shiller cyclically adjusted price-earnings ratio (CAPE) used for prediction of future total returns of the stock market. We treat earnings growth as exogenous. The difference between log wealth and log earnings is modeled as an autoregression of order 1 with linear trend 4.6\% and Gaussian innovations. Detrending gives us a new valuation measure. Our results disprove the Efficient Market Hypothesis.  Therefore, long-run total returns equal long-run earnings growth plus 4.6\%. We apply results to retirement planning. A withdrawal process governs how a retired capital owner withdraws a certain fraction of wealth annually. We study the long-term behavior of such processes.
\end{abstract}

\subjclass[2020]{60F05, 60G10, 60J20, 62J05, 62M10, 62P05, 62P20, 91B84}

\keywords{Autoregression, total returns, earnings growth, ergodic process. \newline {\it Acknowledgements.} The author is thankful to UNR students Akram Reshad, Taran Grove, Erick Luerken, Tran Nhat, and Michael Reyes for useful discussion. The author thanks the Department of Mathematics and Statistics at the University of Nevada, Reno, for welcoming atmosphere for research. The author thanks referees and editors for useful comments.  The author received no funding and declares no conflict of interest.}

\maketitle

\thispagestyle{empty}

\section{Introduction} 

\subsection{Index, earnings, and dividends} 
\begin{figure}[t]
\subfloat[Index]{\includegraphics[width = 8cm]{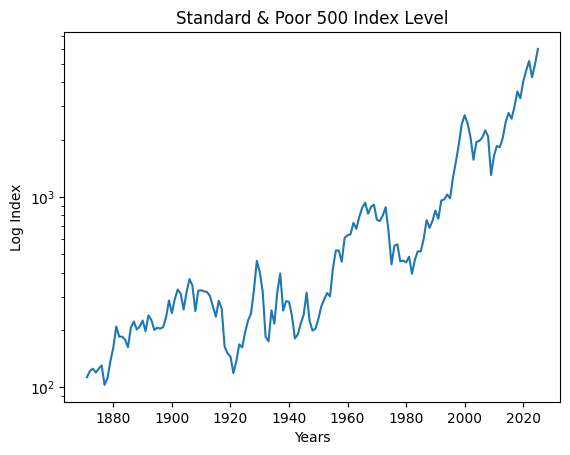}}
\subfloat[Wealth, 1871 = 1\$]{\includegraphics[width = 8cm]{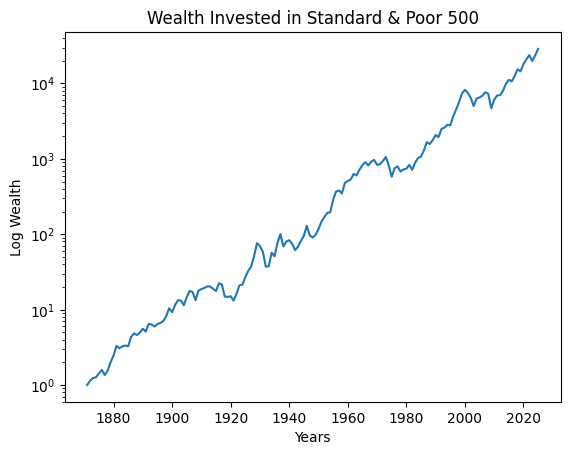}}
\caption{Inflation-adjusted index and wealth 1871--2025. The index grew two orders of magnitude in 154 years, But wealth grew much more: 1\$ invested in 1871 becomes $\approx 20000\$$ in 2025.}
\label{fig:index}
\end{figure}

Forecasting future stock market returns is a major area of study in financial mathematics. We measure stock market by the Standard \& Poor (S\&P) 500 and its predecessors. Robert Shiller collected USA annual stock market data from 1871, which we use in this article. The index (real, that is, inflation-adjusted) is in \textsc{Figure~\ref{fig:index} (A).} The wealth (real) is in \textsc{Figure~\ref{fig:index} (B).} It grows much faster than the index, because it includes reinvested dividends. Dividends are part of earnings (net income) paid by companies to shareholders (usually quarterly or annually). The rest of earnings are retained by companies for other purposes. The combined dividends/earnings of all S\&P 500 companies (including the ones with negative earnings or zero dividends) form S\&P 500 annual dividends/earnings. Returns come from: (a) index increases; (b) dividends payout.  

\begin{figure}[t]
\subfloat[Annual, 1871--2024]{\includegraphics[width = 8cm]{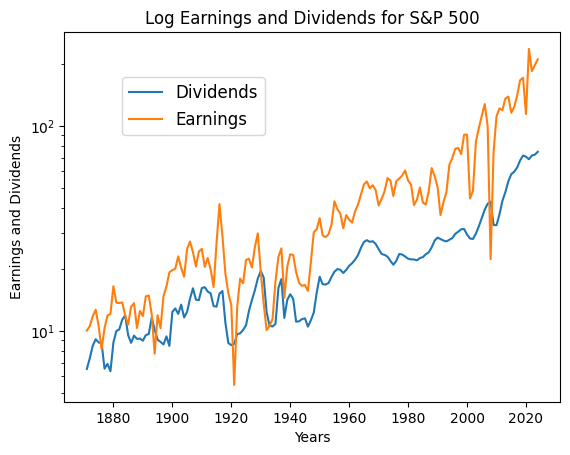}}
\subfloat[5-Year Trailing, 1875--2024]{\includegraphics[width = 8cm]{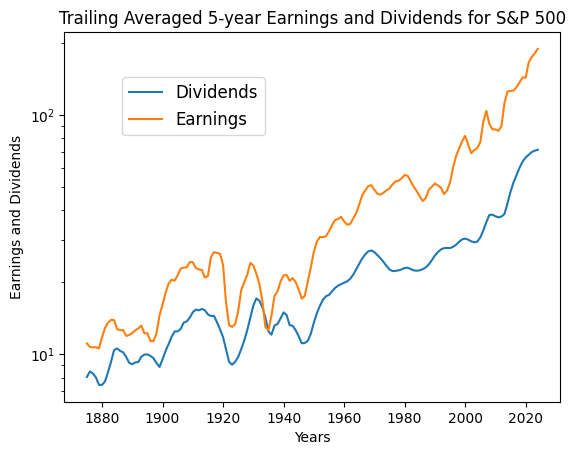}}
\caption{Real earnings and dividends, annual and 5-year trailing averaged}
\label{fig:fundamentals}
\end{figure}

Annual earnings and dividends are in \textsc{Figure~\ref{fig:fundamentals} (A)}. Averaged 5-year earnings and dividends are in \textsc{Figure~\ref{fig:fundamentals} (B)}. One can see that in both graphs, in almost all years, earnings exceed dividends. That is, the upper graph is earnings, and the lower graph is dividends. However, earnings of some companies can be negative. This creates a problem called {\it aggregation bias:} Adding negative earnings of these companies to positive earnings of other more successful companies understates the total. Owning companies with negative earnings does not diminish your claims on positive earnings of other companies.

\subsection{P/E ratios} There is a long tradition of using different versions of a price-to-earnings (P/E) ratio. A stock market or portfolio with high P/E ratio is deemed overpriced, with bearish growth prospects. Conversely, a stock market of portfolio with low P/E ratio is considered attractively underpriced. Robert Shiller designed a {\it cyclically adjusted price-earnings} (CAPE) ratio which takes 10-year averaged inflation-adjusted earnings in the denominator,  \cite{Shiller1998, ShillerBook}. See a discussion in \cite[Chapter 11]{SiegelBook}. A related concept is {\it value investing} which applies the same concept to individual stocks, see the classic article \cite{FF1993} and \cite[Chapter 12]{SiegelBook}.  Instead of surveying all existing literature,  we refer the reader to the books \cite{RW12, ShillerBook, SiegelBook} cited above, as well as \cite{Arnott, JPMorgan, Acct, Ural, PE}. 

We consider both annual S\&P 500 earnings and dividends data, as well as 5-year trailing averages of each. We average to smooth out business cycle fluctuations. During recessions, earnings (and, to a lesser extent, dividends) temporarily plummet, but the impact upon long-term earnings prospects is small. Thus P/E ratio can increase during recessions, but this does not mean that the stock market is overpriced. The CAPE ratio does not have this problem, because it uses averaged earnings. We mentioned aggregation bias above. It is most pronounced during recent recessions: the dot-com crash and the financial crisis. (This remark does not apply to dividends, because they are always nonnegative.) However, we have only this aggregated data. This is an additional reason to average trailing data.

We can also use dividends instead of earnings to compute these ratios. As mentioned earlier, some earnings are paid as dividends, others are retained by companies for other uses. Thus dividends are usually less than earnings. On the other hand, earnings can be negative, but dividends are always nonnegative. 

\begin{figure}[t]
\subfloat[Annual, 1872--2024]{\includegraphics[width = 8cm]{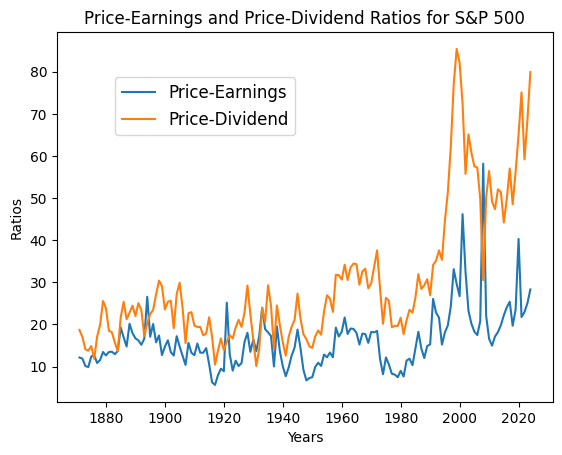}}
\subfloat[Cyclically adjusted, 1876--2024]{\includegraphics[width = 8cm]{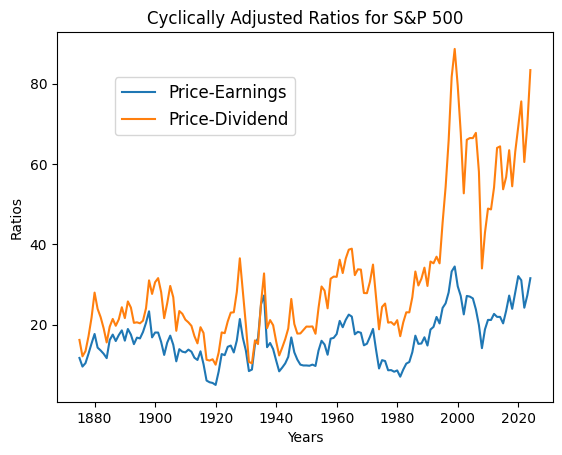}}
\caption{Price-earnings and price-dividend ratios. Left: using annual earnings or dividends. Right: using 5-year trailing averaged earnings or dividends}
\label{fig:ratios}
\end{figure}

This CAPE ratio has significant predictive power for future total real stock market returns (including dividends and price changes). We use the 5-year instead of the 10-year version here, to have slightly more data points (150 instead of 145). However, these trailing versions are not very different. The price-earnings and price-dividend ratios are shown in \textsc{Figure~\ref{fig:ratios}}: the classic versions (without averaging) in \textsc{Figure~\ref{fig:ratios} (A)} and the versions with 5-year averaging in \textsc{Figure~\ref{fig:ratios} (B).} Since earnings exceed dividends, as mentioned before, in both graphs, the price-dividend ratios exceed the price-earnings ratios for the same year. 

\subsection{The new valuation measure} There is evidence that companies prefer to distribute earnings to shareholders via buybacks, rather than dividends. This pushes price upward and increases these P/E ratios, but does not necessarily affect total returns (which combine both price changes and dividend payouts). We consider another valuation measure. 

Let $V(t)$ be the real wealth in (January of) year $t+1$ given $V(0) = 1$. Then $Q(t) = \ln(V(t)/V(t-1))$ are total real returns (including dividends) in year $t$. Averaged annual returns 1871--2024 are $6.7\%$. Let $F(t)$ be the fundamentals (earnings or dividends, annual or trailing 5-year) in year $t$.  Then $G(t) = \ln(F(t)/F(t-1))$ is the (logarithmic) growth in fundamentals. Averaged annual growth is approximately $2\%$. Next, consider  
\begin{equation}
\label{eq:diff-intro}
U(t) = \ln V(t) - \ln F(t).
\end{equation}
It grows annually at rate $c \approx 6.7\% - 2\%$, and can be modeled as an autoregression of order $1$ with linear trend $ct$. Namely, first detrend $U$ by subtracting
\begin{equation}
\label{eq:detrend-intro}
H(t)  = U(t) - ct.
\end{equation}
Then we model $H$ from~\eqref{eq:detrend-intro} as an autoregression of order 1 with Gaussian innovations 
$\varepsilon$:
\begin{equation}
\label{eq:AR-intro}
H(t) - h = \beta(H(t-1) - h) + \varepsilon(t),\quad \varepsilon(t) \sim \mathcal N(0, \sigma^2)\quad \mbox{i.i.d.}
\end{equation}
Here, $\beta \in (0, 1)$; a classic Student $t$-test (which we can apply because the innovations are Gaussian) shows that $\beta \ne 1$ significantly. Therefore, $H$ from~\eqref{eq:detrend-intro} is not a random walk. This disproves the Efficient Market Hypothesis, because we have predictability in returns via~\eqref{eq:diff-intro}. 

We use $H$ as the new valuation measure. These two valuation measures (CAPE and $H$) are close until the last 20 years, then they diverge. You can see two measures on \textsc{Figure~\ref{fig:compare}.} As of January 2025, the CAPE shows that the market is overpriced, but $H$ shows that it is underpriced. As mentioned earlier, this is due to recently changed dividend payout policy. They give opposite results: The new measure indicates that currently (as of January 2025) the U.S. stock market is undervalued and future returns are higher than historical averages. The old measure indicates the opposite. 

This regression~\eqref{eq:AR-intro} has $R^2 = 8\%$. A statistican could raise questions about relevance and importance of this regression. However, it is normal to get such low $R^2$ for financial prediction. The stock index movement is very close to a random walk. Still, here we got statistically significant results, according to Student $t$-tests. We stress that our research holds only in the long run. The valuation measure shows long-term potential. But the next year or two might as well be different. Consider year 1999: the stock market was widely overpriced according to any measure, Shiller CAPE or our valuation measure. However, in 2000 it still zoomed upwards.

\subsection{Retirement} A retiree invests in the S\&P 500 index and withdraws annually a certain fraction of wealth. The resulting wealth process grows slower than the original wealth process. A classic example of a withdrawal strategy is the 4\% rule, which exists in two versions: {\it proportional} (4\% of current wealth) and {\it constant} (4\% of inflation-adjusted initial wealth). In this article, we study the proportional versions. A withdrawal strategy is called {\it sustainable} if the wealth process almost surely converges to infinity. In Section 4, we state and prove sufficient conditions for the withdrawal strategy to be sustainable or not sustainable. In particular, the 4\% proportional rule is, in fact, sustainable. 

\subsection{Organization of the article} We perform this statistical analysis for~\eqref{eq:AR-intro} assuming unknown $c$ in Section 2. We consider four versions of fundamentals $F$: earnings or dividends, each either annual or 5-year trailing averaged versions. Next, in Section 3, we state and prove long-term stability results, assuming long-term stability of earnings or dividends growth, and using the long-term stability for the autoregression of order 1 in~\eqref{eq:AR-intro}: ergodicity in~\ref{thm:ergodic}, Strong Law of Large Numbers in Theorem~\ref{thm:SLLN}, and the Central Limit Theorem in Theorem~\ref{thm:CLT}. In Section 4, we consider {\it withdrawal strategies:} A retiree withdraws certain amounts each year from wealth invested in S\&P 500. We state Theorem~\ref{thm:withdrawal} for proportional withdrawal strategies (fractions of current wealth). Section 5 is devoted to discussions: detailed explanations of economic and financial sense behind this statistical analysis. Section 6 is devoted to conclusions and suggestions for future research. The Appendix contains proofs of all results. The \texttt{GitHub} repository \texttt{asarantsev/IDY} contains our Python code and data. We also include there all figures generated by this code, including figures not included in this article. 

\section{Statistical Analysis}

\subsection{Basic analysis} 
\begin{figure}[t]
\subfloat[$\varepsilon(t)$ QQ]{\includegraphics[width = 5.5cm]{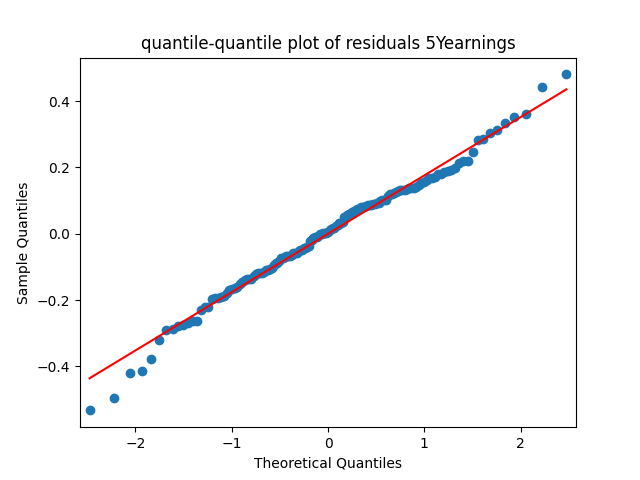}}
\subfloat[$\varepsilon(t)$ ACF]{\includegraphics[width = 5.5cm]{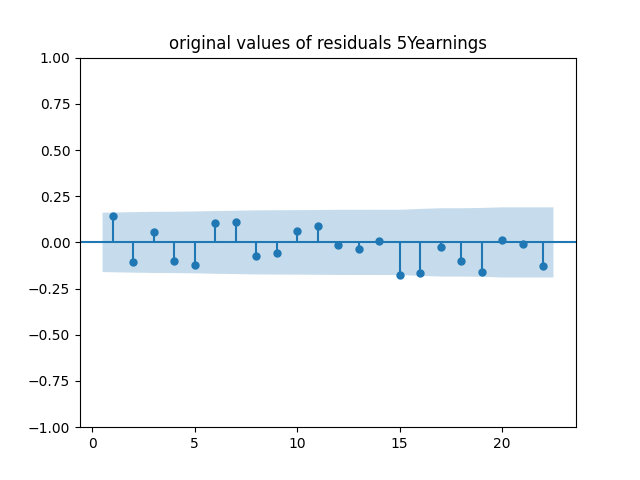}}
\subfloat[$|\varepsilon(t)|$ ACF]{\includegraphics[width = 5.5cm]{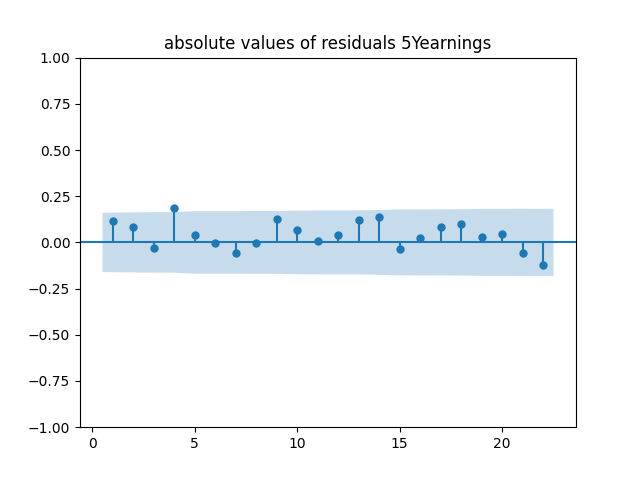}}
\caption{The quantile-quantile plot for the residuals vs the normal distribution; ACF for the residuals and for their absolute values. We take 5-year trailing averaged earnings as a measure of fundamentals. These three plots clearly show $\varepsilon(t)$ can be well modeled by i.i.d. Gaussian.}
\label{fig:res}
\end{figure}

Combine~\eqref{eq:diff-intro},~\eqref{eq:detrend-intro},~\eqref{eq:AR-intro} into one linear regression: 
\begin{equation}
\label{eq:main-reg}
\ln V(t) - \ln F(t) - h - ct = b(\ln V(t-1) - \ln F(t-1) - h - c(t-1)) + \varepsilon(t).
\end{equation}
We can rewrite~\eqref{eq:main-reg} as follows:
\begin{equation}
\label{eq:final-reg}
Q(t) - G(t) = \alpha + \beta t -\gamma(\ln V(t-1) - \ln F(t-1)) + \varepsilon(t),
\end{equation}
where the old and new coefficients are related as follows:
\begin{equation}
\label{eq:new-coeff}
b = 1 - \gamma,\quad c = \frac{\beta}{\gamma},\quad h = \frac{\alpha - c}{\gamma}.
\end{equation}
First, fit the regression~\eqref{eq:final-reg} for 5-year averaged trailing real earnings. See \textsc{Table~\ref{table:reg-results}} for results. As for the regression residuals $\varepsilon(t)$, in \textsc{Figure~\ref{fig:res}} we see their quantile-quantile (QQ) plot vs the normal distribution, and the autocorrelation function (ACF) for $\varepsilon(t)$ and $|\varepsilon(t)|$. Based on this analysis (and the statistical tests found in \textsc{Table}~\ref{table:overall-results}), it is reasonable to conclude that $\varepsilon(t)$ are i.i.d. Gaussian. 

\begin{table}
\begin{tabular}{|c|c|c|c|}
\hline
coefficient & point estimate & $p$-value & [2.5\%, 97.5\%] CI \\
\hline
$\alpha$ & 0.1 & 0.2\% & [0.038, 0.162] \\
\hline
$\beta$ & 0.0071 & 0.1\% & [0.003, 0.011] \\
\hline 
$\gamma$ & 0.1569 & 0.1\% & [0.069, 0.245] \\
\hline
\end{tabular}
\vspace{0.2cm}
\caption{Regression Results~\eqref{eq:final-reg} when fundamentals are 5-year trailing averaged earnings. We see all coefficients are significantly different from zero.}
\label{table:reg-results}
\end{table}

In \textsc{Table~\ref{table:reg-results}}, all coefficients are significantly different from zero. Of particular importance is the coefficient $\gamma$, which is responsible for mean reversion. The parameter $b$ is between $0$ and $1$. This implies that the autoregression~\eqref{eq:final-reg} is mean-reverting, that is, has long-term stability. Plugging point estimates for $\alpha, \beta, \gamma$ from \textsc{Table~\ref{table:reg-results}} into~\eqref{eq:new-coeff}, we get:
\begin{equation}
\label{eq:new-est}
b = 0.84,\quad c = 4.6\%,\quad h = 0.35.
\end{equation}
These point estimates in~\eqref{eq:new-est} are important. Below, we discuss each of these parameters. 

\subsection{Economic meaning of the autoregression coefficients} The parameter $b$ is responsible for mean reversion: Since $b \in (0, 1)$, this  autoregression is indeed mean-reverting, not a random walk. As discussed, we reject the unit root hypothesis, which in this context means $b = \pm 1$. The closer $b$ to zero, the stronger the mean reversion is. 

The long-run average of the implied dividend yield $\Delta$ is $c$. This can be viewed as the long-term value of the second component of returns. As discussed earlier, this number matches the return on the capital in traditional societies with zero economic and earnings growth. Together with earnings growth (with long-term average $g = 1.9\%$), this gives us total real returns, with long-term average $c + g = 4.6\% + 1.9\% = 6.5\%$; see Theorem~\ref{thm:SLLN}.  

Finally, $h$ is the long-term average of the new valuation measure. Comparing this measure with $h$, one can find whether the stock market is over- or underpriced. The current (as of January 2025) value of $H$ is equal to $0.196 < h = 0.36$. Thus the current stock market is underpriced. Future total real returns, given the same real earnings growth as in the last 154 years, are higher (than the overall long-term average $c + g$). 

\subsection{Four versions of fundamentals} In \textsc{Table}~\ref{table:overall-results}, we present the results of regression analysis~\eqref{eq:final-reg} for four versions of the fundamentals: earnings/dividends, each either annual or 5-year trailing averaged. 

\begin{table}
\begin{tabular}{|c|c|c|c|c|}
\hline
\hline
Statistics & 1E & 1D & 5E & 5D \\
\hline
\hline
Mean of growth & 1.99\% & 1.59\% & 1.9\% & 1.47\% \\
\hline
Stdev of growth & 28\% & 11\% & 8\% & 5\% \\
\hline
Stdev of $\varepsilon$ & 25\% & 19\% & 18\% & 17\% \\
\hline
Regression $R^2$ & 16.4\% & 12.6\% & 8.0\% & 9.5\% \\
\hline
\hline
Long-term valuation $h$ & 0.32 & 0.11 & 0.35 & 0.23 \\
\hline
Mean-reversion $b$ & 0.67 & 0.75 & 0.84 & 0.81 \\
\hline
Linear trend $c$ & 4.6\% & 5\% & 4.6\% & 5\% \\
\hline
Coefficient $\gamma$ & 0.33 & 0.25 & 0.16 & 0.19 \\
\hline
Student $T(\gamma)$ & 5.4 & 3.4 & 4.5 & 3.8 \\
\hline
Jan 2025 valuation $H$ & 0.062 & -0.022 & 0.196 & 0.123 \\
\hline
Pearson correlation & -73\% & -42\% & -28\% & -26\% \\
\hline
Spearman correlation & -60\% & -40\% & -20\% & -19\% \\
\hline
\hline
SW $p$ & 0.1\% & 29\% & 39\% & 2.3\% \\
\hline 
JB $p$ & $< 0.1\%$ & 25\% & 28\% & 0.3\% \\
\hline
O5 $p$ & 82\% & 4.2\% & 10\% & 3.3\% \\
\hline
O10 $p$ & 37\% & 0.3\% & 14\% & 4.7\% \\
\hline
A5 $p$ & 24\% & 5.8\% & 12\% & 11\% \\
\hline
A10 $p$ & 25\% & 10.5\% & 25\% & 15\% \\
\hline
\hline
\end{tabular}
\vspace{0.2cm}
\caption{Results of regression~\eqref{eq:final-reg} for four versions of fundamentals: 1-year earnings (1E) or dividends (1D), 5-year averaged trailing earnings (5E) or dividends (5D). Only for 5-year averaged trailing earnings (5E) we can conclusively accept that residuals are i.i.d. Gaussian.}
\label{table:overall-results}
\end{table}

In the first part, we provide basic analysis. First, we give the empirical mean and standard deviation of growth terms $G$, and the empirical standard deviation of regression residuals $\varepsilon$. Then we provide the $R^2$ for this regression, and the values of $b, c, h$. Next, we provide the value of the coefficient $\gamma$ from~\eqref{eq:final-reg} and the absolute value of the Student $t$-statistics $T(\gamma)$. 

We also present the current value of the valuation measure (as of January 2025). Comparing it with the long-term value $h$, we find whether the market is over- or undervalued. In the second part, we perform analysis of regresson residuals $\varepsilon(t)$. The $p$-values with indices SW, JB are for Shapiro-Wilk and Jarque-Bera normality tests for $\varepsilon(t)$. The $p$-values with indices O5, O10 are for Ljung-Box test for $\varepsilon(t)$ with 5 and 10 lags, respectively. The $p$-values with indices A5, A10 are for Ljung-Box test for $|\varepsilon(t)|$ with 5 and 10 lags, respectively. We see that some $p$-values are large, others are much smaller than the classic 5\% threshold. But only for the 5-year trailing averaged earnings we can strongly conclude that residuals $\varepsilon(t)$ of~\eqref{eq:final-reg} are i.i.d. Gaussian.

Finally, we compute Pearson and Spearman correlations between $G(t)$ and $\varepsilon(t)$. All of them are negative and significantly different from zero. The Python code shows $p$-values, and all of them are much smaller than $5\%$. 

Real earnings/dividends growth terms $G(t)$ (in each of the four versions) are not white noise (judging by the ACF) and not Gaussian (judging by the Shapiro-Wilk and Jarque-Bera normality tests). This is why we compute Spearman correlations between $G(t)$ and $\varepsilon(t)$. In \textsc{Table~\ref{table:overall-results}}, we conclude that $R^2$ is the best for one-year versions. 

This is important: we model $U$ as an autoregression of order 1 with linear trend (which requires residuals to be white noise), and apply classic Student tests (which requires normality of residuals). 

A low value of $R^2$ in all these cases is not a judgement of the quality of regression analysis. This is simply a reflection of the observation that stock index returns have a lot of randomness and little predictability. However, for all these regressions the coefficient $\gamma$ is significant: The Student (Gosset) $t$-statistics is always greater than $3$. This points to mean reversion and returns predictability, and disproves the Efficient Market Hypothesis. 

Additionally, in each case, the current valuation measure is less than the long-term average $h$. Thus the market is undervalued under each of these analyses. The long-run returns are  lower than the theoretical average $c + g$. 

Standard deviation of total returns is 17\%. If we treat fundamentals growth as exogenous, we must choose the version with the smallest standard deviation. The less this standard deviation is, the more the variance reduction is. This version is the 5D with standard deviation 5\%. The next smallest one is 5E with standard deviation 8\%. 


\subsection{Why the new valuation measure is superior to Shiller CAPE} The classic way to use CAPE for index returns prediction is to perform a simple linear regression of next year's total real returns upon the CAPE value for this year. However, this results in a failure on multiple levels. The correlation is $R = -12\%$ so $R^2 = 1.5\%$, much lower than $R^2 = 8\%$ for the new valuation measure. The $p$-value for the slope is $13\%$, thus we fail to reject the null hypothesis (that the CAPE and next year's returns are uncorrelated). Finally, and most importantly, the residuals fail normality tests: Shapiro-Wilk and Jarque-Bera tests give us $p = 0.4\%$ and $p = 0.5\%$. See also \textsc{Figure}~\ref{fig:compare} for comparison of the two measures. We take the logarithm of CAPE to make the same scale, and normalize so that both measures are equal to zero in 1876. 

\section{Model Statement and Theoretical Results} 

\subsection{Background definitions} In this section, we create a time series model based on the statistical analysis in the previous section. We operate on a filtered probability space $(\Omega, \mathcal F, (\mathcal F_0, \mathcal F_1, \ldots), \mathbb P)$. We state the following well-known definitions from Stochastic Processes theory.

\begin{definition} A real-valued discrete-time stochastic process $X = (X_0, X_1, \ldots)$ is called {\it adapted} with respect to the given discrete-time filtration if $X_t$ is $\mathcal F_t$-measurable. The process $X$ has a {\it stationary distribution} $\pi$ if $X_t \sim \pi$ for all $t = 0, 1, 2, \ldots$
\label{defn:general}
\end{definition}

For probability measures $P$ and $P'$ on $\mathbb R^2$, define the {\it total variation distance}: 
\begin{equation}
\label{eq:TV}
d_{\mathrm{TV}}(P, P') = \sup\limits_{A \subseteq \mathbb R^2}|P(A) - P'(A)|,
\end{equation}
and the {\it Wasserstein distance of order} $p \ge 1$: 
\begin{equation}
\label{eq:W-p}
\mathcal W_p(P, P') = \inf\limits_{X \sim P, \, Y \sim P'}\mathbb E\left[|X - Y|^p\right].
\end{equation}
In~\eqref{eq:W-p}, the inf is taken over all {\it couplings:} pairs of random variables $(X, Y)$ such that $X \sim P$ and $Y \sim P'$. Both these functions are indeed distances:~\eqref{eq:TV} is on the set $\mathcal P$ of all probability measures on $\mathbb R^2$, and~\eqref{eq:W-p} is on the set of all $P \in \mathcal P$ with finite $p$th moment. Finally, a discrete-time Markov process $X = (X(0), X(1), \ldots)$ on the state space $\mathbb R^2$ has a {\it stationary distribution} $\Pi$ (a probability measure on $\mathbb R^2$) if $X(0) \sim \Pi$ implies $X(1) \sim \Pi$ (hence $X(t) \sim \Pi$ for all $t$).

\subsection{Model statement} We gather and reverse-engineer the equations from the previous section. Starting from 
$\varepsilon(t)$ and $F$, we define $G$, $H$, and $Q$. 

\begin{definition} Take a positive adapted stochastic process $F = (F(0), F(1), \ldots)$ which we call {\it the fundamental process}. Take another adapted white noise sequence 
$\varepsilon(t) \sim \mathcal N(0, \sigma^2)$ for $t = 1, 2, \ldots$ Here, $\varepsilon(t)$ is independent of $\mathcal F_{t-1}$. Fix constants $b \in (0, 1), c > 0$, and {\it long-term valuation measure} $h$. Define {\it valuation measure} $H$ using the equation~\eqref{eq:main-reg}, with $H(0)$ an $\mathcal F_0$-measurable random variable. {\it Fundamental growth} $G(t)$ for year $t$ is defined by
\begin{equation}
\label{eq:growth-def}
G(t) = \ln\frac{F(t)}{F(t-1)},\, t = 1, 2, \ldots
\end{equation}
and the {\it total real returns} $Q(t)$ for year $t$ are defined as
\begin{equation}
\label{eq:model}
Q(t) = H(t) - H(t-1) + c + G(t),\, t = 1, 2, \ldots
\end{equation}
\label{defn:process}
\end{definition}

\begin{definition} {\it Average annual total real returns} for $T$ years are defined as 
\begin{equation}
\label{eq:def-avg}
\overline{Q}(T) = \frac1T\left[Q(1) + \ldots + Q(T)\right].
\end{equation}
{\it Real wealth} by year $T$ (starting from wealth $1$ at year $0$) is 
\begin{equation}
\label{eq:def-wealth}
V(T) = \exp(Q(1) + \ldots + Q(T)) = \exp(T\overline{Q}(T)).
\end{equation}
\label{defn:avg}
\end{definition}

We do {\it not} require $\varepsilon$ and $G$ to be independent or uncorrelated. Together, Definitions~\ref{defn:process} and~\ref{defn:avg} form a discrete-time model. This is the main model of this article. In the rest of this section, we analyze this model. 

\subsection{Assumptions} We impose a few assumptions on the long-term behavior. We believe these assumptions seem reasonable for real-life data. Indeed, earnings or dividends seldom grow or fall more than twice (or any other reasonable bound) per year, so we can assume this growth has finite moments of any order. Also, growth a long time ago does not influence growth now, so one can safely assume strong mixing conditions. They imply the SLLN and the CLT. We also require continuity of the transition density, since this is important for later stability proofs. We think this continuity assumption is reasonable and is usually easily satisfied. There exist a lot of continuous distributions supported on the entire space, for example multivariate normal or Laplace, or their generalizations. We emphasize we do {\it not} require that these growth terms are i.i.d. Growth of fundamentals in 2023 might depend upon growth in 2022. 

\begin{asmp} The process $G$ satisfies the SLLN: almost surely, 
$$
\overline{G}(T) := \frac{G(1) + \ldots + G(T)}{T} = \frac1T\ln\frac{F(T)}{F(0)} \to g,\quad T \to \infty.
$$
\label{asmp:LLN}
\end{asmp}

The next assumption is the Central Limit Theorem (CLT) for the real earnings growth.

\begin{asmp} The process $G$ satisfies $\sqrt{T}(\overline{G}(T) - g) \to \mathcal N(0, \rho^2)$ in law as $T \to \infty$. 
\label{asmp:CLT}
\end{asmp}

Next, we state a Markovian assumption: that real earnings growth term $G(t)$ for time $t$ depends not on the entire history $G(1), \ldots, G(t-1)$ and $\varepsilon(1), \ldots, \varepsilon(t)$, but instead only on $G(t-1)$ and $\varepsilon(t)$.

\begin{asmp}
For a {\it transition density} $\mathbf{p} : \mathbb R^3 \to (0, \infty)$ continuous in $(x, y)$, 
$$
\mathbb P(G(t) \in A\mid G(t-1) = x, \varepsilon(t) = y) = \int_A\mathbf{p}(x, y, z)\,\mathrm{d}z.
$$
\label{asmp:Markov}
\end{asmp}

Finally, we state an assumption of bounded moments, necessary for convergence proofs. We already explained above why this assumption is reasonable.

\begin{asmp} $\sup_{t \ge 1}\mathbb E[G^{p}(t)] < \infty$ and $\mathbb E[H(0)|^p < \infty$ for some $p > 1$. 
\label{asmp:bdd}
\end{asmp}

\subsection{Main results for total real returns} The proofs are in the Appendix. First, we state a technical lemma.

\begin{lemma} Average annual total real returns for $T$ years are given by
\begin{equation}
\label{eq:mean-return}
\overline{Q}(T) = c + \frac{H(T) - H(0)}T + \overline{G}(T),
\end{equation}
and wealth at time $T$ is given by 
\begin{equation}
\label{eq:total-wealth}
V(T) = \exp\left[cT + H(T) - H(0)\right]\frac{F(T)}{F(0)}.
\end{equation}
\label{lemma:main}
\end{lemma}

Our results for the total real returns are: the Strong Law of Large Numbers (SLLN), Markovian property, ergodicity, and the CLT. 

\begin{theorem} Under Assumption~\ref{asmp:LLN}, we have the following asymptotic results: as $T \to \infty$, 
\begin{align}
\label{eq:limiting-delta}
\frac1T(\ln V(t) - \ln F(t)) \to c\quad \mbox{almost surely.}
\end{align}
Average total real returns satisfy the convergence statement:
\begin{equation}
\label{eq:R-LLN}
\overline{Q}(T) \to c + g\quad \mbox{almost surely as}\quad T \to \infty.
\end{equation}
\label{thm:SLLN}
\end{theorem}

\begin{theorem}
\label{thm:Markov}
Under Assumption~\ref{asmp:Markov}, the process $(G, H)$ is Markov: For $A \subseteq \mathbb R^2$ and $t$,
\begin{align*}
\mathbb P&\left((G(t), H(t)) \in A\mid G(t-1), H(t-1)\right) \\ & = \mathbb P\left((G(t), H(t)) \in A\mid H(0), G(1), H(1), \ldots, G(t-1), H(t-1)\right). 
\end{align*}
\end{theorem}

\begin{theorem}
\label{thm:ergodic}
Under Assumption~\ref{asmp:Markov} and~\ref{asmp:bdd}, 

(a) the process $(G, H)$ has a unique stationary distribution $\Pi$ on $\mathbb R^2$ with positive Lebesgue density on $\mathbb R^2$, with finite $p$th moments, and a Gaussian marginal for $H$: $\Pi_H = \mathcal N(h, \sigma^2_{\Pi})$, where 
$\sigma^2_{\Pi} = \sigma^2/(1 - b^2)$;

(b) regardless of the initial distribution, the distribution of $(G(t), H(t))$ converges to $\Pi$ as $t \to \infty$ in the total variation and in the Wasserstein distance of order $p$.  
\end{theorem}

Finally, we state the CLT for total real returns: In the long run, wealth is a lognormal random variable. This follows from the CLT assumption for the real earnings growth series. 

\begin{theorem}
Under Assumptions~\ref{asmp:LLN}, ~\ref{asmp:CLT}, we have weak convergence
$$
\sqrt{T}(\overline{Q}(T) - c - g) \to \mathcal N(0, \rho^2),\quad T \to \infty.
$$
\label{thm:CLT}
\end{theorem}

\section{Retirement Planning} 

A major problem in finance is {\it retirement planning}: How much a retiree can withdraw per year to keep income more or less stable, and still preserve wealth? Wealth funds and university endowments face a related problem; the difference is that, unlike the retiree, they wish to preserve the wealth forever, not for 20--30 years. The classic withdrawal rule is 4\%:

\begin{itemize}
\item {\it Constant:} Withdraw inflation-adjusted 4\% of initial wealth; 
\item {\it Proportional:} Withdraw 4\% of current wealth.
\end{itemize}

The first rule ensures constant stream of payments, but could deplete wealth up to zero ({\it ruin}, or {\it bankruptcy}). A financial planner must ensure that the ruin probability is small. The second rule ensures that there is no ruin, but income could fluctuate (when markets crash, income decreases). In this article, we consider only the proportional rule. We generalize it for the case of changing (and possibly random) proportional withdrawal rate. 

\begin{definition} A {\it withdrawal process} $W = (W(0), W(1), \ldots)$ is a discrete-time stochastic process with values in $(0, 1)$. Its corresponding  {\it wealth process} $V_W$ is defined as
$$
V_W(t) = V_W(t-1)e^{Q(t)}(1 - W(t)),\, t = 1, 2, \ldots;\quad V_W(0) = 1.
$$
\end{definition}

For example, if total real return $Q(t) = 7\%$, and withdrawal rate $W(t) = 4\%$, then wealth from $t-1$ to $t$ is multiplied by $e^{0.07}$ from total real return and is multiplied by $0.96$ from withdrawal. Thus the overall multiplication factor is $e^{0.07}\cdot 0.96 \approx 1.03$. If withdrawal process is not constant, we define an asymptotic rate: a long-term limit.

\begin{definition} A withdrawal process $W$ has {\it asymptotic rate} $w$ if it satisfies
$$
\frac1T\sum\limits_{t=1}^TW(t) \to w\quad \mbox{a.s. as}\quad T \to \infty.
$$
\end{definition}

For any withdrawal process, $V_W(t) > 0$ almost surely for all $t$. But still, $V_W(t)$ could decrease to small values, which is, of course, undesirable. We define a condition under which this does not happen. 

\begin{definition}
A withdrawal process is called {\it sustainable} if $V_W(t) \to \infty$ a.s. as $t \to \infty$. 
\end{definition}

We prove sufficient conditions for sustainability, and other sufficient conditions for non-sustainability. There is a gap between these two results, because $c + g > 1 - e^{-c-g}$, so these are not necessary and sufficient conditions. To close this gap is left for future research. 

\begin{theorem} Under  Assumption~\ref{asmp:LLN}, 

(a) any constant withdrawal process $W(t) = w < 1 - e^{-c-g}$ is sustainable;

(b) any withdrawal process with asymptotic rate $w > c + g$ is not sustainable.
\label{thm:withdrawal}
\end{theorem}

For example, if we take 5-year earnings as fundamentals, then $c + g = 6.4\%$ and therefore $1 - e^{-c-g} = 1 - e^{-0.064} = 0.062$. Thus any withdrawal process with constant proportional rate less than $6.2\%$ in sustainable. For example, the classic 4\% withdrawal rule (withdrawing 4\% of current wealth) is sustainable. 

\subsection{Withdrawal processes depending on earnings growth} One would like explicit formulas for wealth distribution for simple withdrawal processes, for example $W(t) \equiv w$. However, we need to model real earnings growth time series, but we do not impose any assumptions on it. Thus we cannot find such explicit formulas. However, we can create a special withdrawal process dependent on $G(t)$ with wealth process independent of $G(t)$. 

\begin{asmp}
There exists a constant $u > 0$ such that $G(t) > -u$ almost surely for all $t$.
\label{asmp:growth}
\end{asmp}

\noindent Under Assumption~\ref{asmp:growth}, we can define the withdrawal process 
\begin{equation}
\label{eq:new-withdraw}
W(t) = 1 - \exp(u + G(t)).
\end{equation}

\begin{lemma} Under Assumption~\ref{asmp:growth}, the process~\eqref{eq:new-withdraw} is sustainable if and only if $c > u$. 
\label{lemma:new-withdraw}
\end{lemma}

\section{Background \& Discussion} 

\begin{figure}[t]
\includegraphics[width = 14cm]{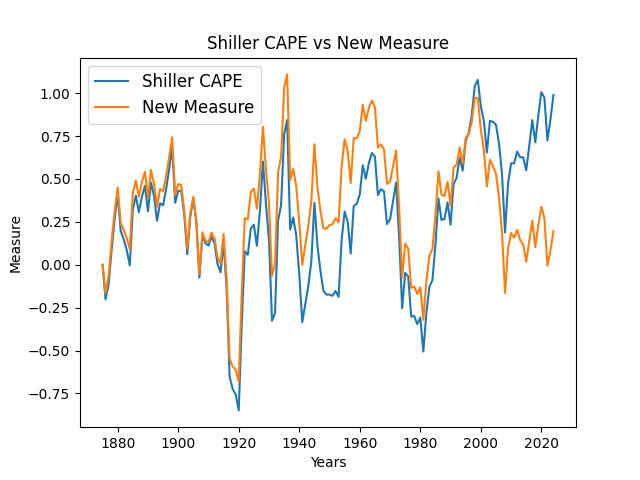}
\caption{The log CAPE and the new valuation measure (normalized so thar both are zero in 1876). We see they are quite close until the last few decades.}
\label{fig:compare}
\end{figure}

\subsection{Sources of stock market returns} The long-run total returns of a stock portfolio, or an individual stock, or the entire stock market, are composed of three parts: dividend yield (dividends paid last year divided by the current price of a stock), earnings growth, and changes in the {\it P/E ratio}, or {\it price-to-earnings ratio}. For the market as a whole (measured by the Standard \& Poor 500 or any other index), the P/E ratio is computed as the sum of P/E ratios of individual stocks, weighted by market capitalizations (total market value) of stocks.  or, equivalently, the ratio of the total market value of all these stocks to their total earnings over the last year. To quote the book \cite[pp.324--325]{RW12}:

\begin{quote}
Very long-run returns from common stocks are driven by two critical factors: the dividend yield at the time of purchase, and the future growth rate of earnings and dividends. In principle, for the buyer who holds his or her stock forever, a share of common stock is worth the present, or discounted value of its stream of future dividends. Recall that this discounting reflects the fact that a dollar received tomorrow is worth more than a dollar received today. [...] The discounted value of this stream of dividends (or funds returned to shareholders through stock buybacks) can be shown to produce a very simple formula for the long-run total return for either an individual stock or the market as a whole: Long-run equity return = initial dividend yield + growth rate. [...] Over shorter periods, such as a year or even several years, a third factor is critical in determining returns. This factor is the change in valuation relationships --- specifically, the change in the price-dividend or price-earnings multiple. 
\end{quote}

In other words, dividend yield and earnings growth produce {\it fundamental return}, driven by {\it fundamentals} (earnings and dividends), while changes in the P/E ratio produce {\it speculative return}, driven by market emotions. A classic result in financial theory is that if a P/E ratio of the whole market (measured by S\&P or any other comprehensive index) is high relative to the long-term average (which is approximately 16), then the market is overpriced and will deliver low returns in the near future. 

The difference between total returns and earnings growth fluctuates as an autoregression of order 1 around its mean value, which is approximately 4--5\%. Consider a stagnant economy (with no population growth and no scientific-technological progress and thus no earnings growth). Then the total return fluctuates around 4--5\%. In his classic book, Thomas Piketty observes, see \cite[p.206]{Capital}:

\begin{quote}
\ldots traditional rate of conversion from capital to rent in the 18th and 19th centuries, for the most common and least risky forms of capital (typically land and public debt), was generally on the order of 5\% a year.
\end{quote}

In these agrarian societies, economic growth and therefore earnings growth was almost 0\%. To quote Piketty from \cite[p.353]{Capital}:

\begin{quote}
Economic growth was virtually nil throughout much of human history: cobining demographic and economic growth, we can say that the annual growth rate from antiquity to the 17th century never exceeded 0.1--0.2\% a year. Despite the many historical uncertainties, there is no doubt that the rate of return on capital was always considerably greater than this: the central value observed over the long run is 4--5\% a year. In particular, this was the return on land in most traditional agrarian societies. 
\end{quote}

Thus 4--5\% a year can be thought of as pure return on capital other than growth. The new valuation measure shows stability of this pure return fluctuating around this 4--5\% level. Only earnings growth is left as exogenous. Two other components: dividend yield and valuation changes are modeled as an autoregression of order 1. The new measure describes two out of three components of total returns.

\subsection{Dividends and buybacks} Continuing the quote on \cite[pp.325--326]{RW12}:

\begin{quote}
Many analysts question whether dividends are as relevant now as they were in the past. They argue that firms increasingly prefer distributing their growing earnings to stockholders through stock repurchases rather than dividend payments. Two reasons are offered for such behavior --- one serves shareholders and the other management.
\end{quote}

Companies can distribute earnings to shareholders using buybacks instead of payouts, or reinvest them into business, which raises future earnings and dividends. Whether the {\it payout ratio} (fraction of earnings paid out as dividends) is an important indicator is a hotly debated topic, starting from the classic article \cite{MM61}, which gives a negative answer to this question. To capture their argument informally: Lower dividend yield now leads to more earnings reinvested into business and thus raise future earnings and dividend growth. See the survey \cite{DividendSurvey} for a comprehensive review of literature defending each side of this controversy. Recently, corporate payout and buyback policy was incorporated into the classic P/E research by Robert Shiller, see  \cite{Bunn, Farouk}. Thus we can view annual total real return minus annual real earnings growth as implied dividend yield. Arguably, this implied dividend yield is a more comprehensive measure than actual dividend yield. 

In the classic Shiller CAPE, we compare only prices and earnings without regard to dividends. For the new valuation measure, we compare total returns with earnings growth only, and dividend yield is featured indirectly, through detrending. 

Recall that out of three return components: earnings growth, dividend yield, and changes in the P/E ratio, we labeled the first two {\it fundamental}, and the third one {\it speculative}. Now we merge the second and the third component in this implied dividend yield. To distinguish between its speculative and fundamental parts, we assume that this implied dividend yield fluctuates around the long-term average. In our research, we find that the intrinsic long-run average for this implied dividend yield is $c = 4.6\%$. 

However, in the short run it can deviate from this number. If it significantly exceeded this average for the last few years, then the market can be considered overheated and overpriced. Thus the valuation measure is the cumulative sum of all past annual implied dividend yields, minus true long-term average times the number of years elapsed. This detrending will be included into the regression. Thus we regress next year's implied dividend yield upon last year's detrended valuation measure.

In each two choices of valuation measure $M(t)$: classic CAPE and $H$; for each choice of averaging window $W$, we get the following decomposition of total returns:
\begin{equation}
\label{eq:decomp}
R(t) = \triangle M(t) + \Gamma(t) + \Delta(t),
\end{equation}
where $\Gamma$ is earnings growth rate for trailing 5-year averaged real earnings; $M$ is governed by an autoregression of order 1, with $\triangle M(t) \equiv M(t) - M(t-1)$; and $\Delta(t) = \ln(1 + D(t)/S(t))$ is the logarithmic dividend yield for classic CAPE, and $\Delta(t) = 0$ for the two other measures. Recall the three components of total returns: earnings growth, dividend yield, and valuation changes. In the right-hand side of~\eqref{eq:decomp}, (a) For the classic CAPE, $\triangle M$ is the valuation change, $\Gamma$ is earnings growth, and $\Delta$ is dividend yield. (b) For the measure $H$, $\triangle M$ is the sum of valuation change and dividend yield, and $\Gamma$ is earnings growth.

The original CAPE does not explain stock market returns without an additional term (depending on dividend yield). The two modifications of this model do explain stock market returns as the sum of an autoregression of order 1 and earnings growth. 

Before 1920, earnings growth was 1--2\% while dividend yield was 5\%. After 1970, earnings growth is 4--5\%, dividend yield is 2\%. Dividend yield is chosen by corporate payout policies. If a company chooses to pay much of its earnings as dividends, then dividend yield is large, but retained earnings (used to pay debt, invest in expansion or R\&D, and make stock buybacks) are small; thus earnings growth must be smaller. Conversely, if a company chooses to buy back stock instead of paying dividends, then earnings per share increase faster. Thus actual earnings growth might differ from what one could call fundamental earnings growth, which is growth with non-changing dividend payout/stock buyback policy. Similarly, actual dividend yield might not be representative of fundamental dividend yield. Recall the Modigliani-Miller theorem that states indifference of total returns to dividend policy. 

\subsection{Comparison of 5-year CAPE vs the new valuation measure} 
\begin{figure}
\includegraphics[width = 14cm]{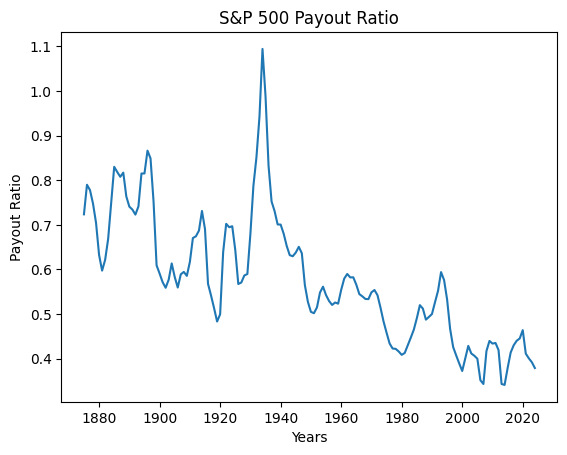}
\caption{Payout Ratio 1876--2024: the ratio of dividends to earnings. For both, we take 5-year averaged trailing rather than the annual quantities, to smooth fluctuations. We see that recently, payout ratio has decreased. This suggests traditional financial ratios might be obsolete.}
\label{fig:payout}
\end{figure}

As of January 2025, the CAPE shows the market is overpriced, but the current valuation measure is lower than its long-term average. In \textsc{Figure~\ref{fig:compare}}, we plot these two valuation measures: log CAPE and $H$. Both are centralized by subtracting their historical averages. We can see that these two measures sharply diverge after the start of the XXI century. This can be attributed to recent reluctance of companies to pay dividends. See \textsc{Figure~\ref{fig:payout}}, showing the {\it payout ratio}: Fraction of earnings paid out as dividends. We take 5-year trailing averages for earnings and dividends to smooth out annual fluctuations. One can clearly see that after 2000, the payout ratio was much lower than the historical average. This change was rather recent (last 20 years). If this aversion to dividends persists for another century, then CAPE might become elevated and no longer representative of the market conditions. Then the new valuation measure will have stronger predictive power of future returns. 

How to explain these variations in the payout ratio? Recent reluctance to pay dividends is partly a consequence of the proliferation of stock options given to executives as part of their compensation package. The latter thus have an incentive to boost the stock price, even at the expense of dividends. Another reason is taxation: Capital gains (increases in stock prices) are often taxed less than dividends. 

During the World War II and the generation after that (1940--1970), in contrast to the new millennium, the new measure is much higher than the log CAPE during 1940--1970. \textsc{Figure~\ref{fig:payout}} explains this as well: Right before this era, during the Great Recession of the 1930s, payout ratio was unusually high compared to the historical average. In fact, for years 1930--1933 annual  dividends paid equaled or exceeded earnings, see \textsc{Figure~\ref{fig:fundamentals}.} In 1934--1939 this ratio dropped but was still greater than 50\%. This strange phenomenon can be explained as follows: During the economic and financial collapse of the 1930s, earnings plummeted. Dividends decreased as well, but not by much. In times of crisis and recession, companies are reluctant to cut dividends, since this hurts investor confidence. Thus it can happen that dividends temporarily exceed earnings. (The same happened in 2008.) 

This anomalously high payout ratio in the 1930s shifted the new valuation measure upward compared to the log CAPE. It stayed this way for three other decades before decreasing below 50\% and staying this way for almost all years since then. Only then the log CAPE caught up with the new valuation measure and then substantially exceeded it. 

\subsection{Adjustment of CAPE for dividends} In the articles \cite{Bunn, Farouk} we get the modification of CAPE version which Shiller called {\it total returns cyclically adjusted price-earnings ratio}, or TR-CAPE. This is computed as follows.

Investing 1\$ at the beginning of data in 1871 and reinvesting dividends, compute total wealth for each year. This wealth grows much faster than the index, as seen in \textsc{Figure~\ref{fig:index}.} Divide this wealth by the index level (adjusted so that 1876 = 1\$) and get the correction coefficient which measures, loosely speaking, wealth produced only by dividends, not by capital gains. We multiply each year's earnings by this correction coefficient. This gives us modified earnings. Average them over the previous 5 years, and divide the wealth by these trailing 5-year modified earnings. We put wealth, not index level, in the numerator, to adjust it for dividends (or, equivalently, for total return) as well. 

Recall our split of market returns into three components: two fundamental (earnings growth and dividend yield) and one speculative (change in valuation). Here we united two fundamental components into one (growth of 5-year trailing total return-adjusted earnings). Thus we can study our speculative component as a stock market valuation measure. This TR-CAPE measure theoretically seems better than the original Shiller CAPE. But, in practice, these measures are very close: Their Pearson correlation is 99\%. The measure $H$ is substantially different from Shiller CAPE; the TR-CAPE is not. Thus we shall not study this modified measure, since it is not an improvement over the classic 5-year CAPE.

\subsection{Future research} It would be interesting to model earnings growth. This will include economic forecasts, since corporate earnings are closely tied to overall economy and tax policy. In addition, a more detailed investigation for withdrawal rates is needed. Alternatively, one can replicate this research for individual stocks and stock portfolios, rather than the overall stock market index. Finally, we can include interest rates in our modeling (short-term Treasury bills, long-term Treasury bonds, and long-term AAA corporate rates), which are also available from 1871 onward. We can use these bond yields in our regression models.  Low bond yields make them less attractive and push stock prices upward, making CAPE higher. Since now we have historically low yields in the USA, this can explain historically high CAPE ratios. In addition, we can make portfolios of stocks (measured by the S\&P index) and bonds (Treasury or AAA corporate bonds). 

\section{Appendix}

\subsection{Proof of Lemma~\ref{lemma:main}} Equation~\eqref{eq:mean-return} follows from~\eqref{eq:def-avg} and the following computations:
\begin{align*}
\overline{Q}(T) & = \frac1{T}\left(\ln V(T) - \ln F(T)  + G(1) + \ldots + G(T)\right) \\ & = c + \frac{H(T)}{T} - \frac{H(0)}{T} + \frac1{T}\left(G(1) + G(2) + \ldots + G(T)\right).
\end{align*}
Equation~\eqref{eq:total-wealth} follows from~\eqref{eq:growth-def} and~\eqref{eq:mean-return}.

\subsection{Proof of Theorem~\ref{thm:SLLN}}  Let us show~\eqref{eq:limiting-delta}. We can rewrite 
$$
\frac1T\left(\ln V(t) - \ln F(t)\right) = \frac1T\left(cT + H(T) - H(0)\right) = c + \frac{H(T) - H(0)}T.
$$
It suffices to show that almost surely,
\begin{equation}
\label{eq:conv-to-0}
\frac{H(T) - H(0)}{T} \to 0,\quad T \to \infty.
\end{equation}
The process $H$ is ergodic in the sense of classic ergodic theory; see for example \cite[Chapter 6]{Stroock}. The mean of the stationary distribution for $H$ is equal to $c$. By the classic Birkhoff ergodic theorem, see \cite[Theorem 6.2.7]{Stroock}, we have almost sure convergence
\begin{equation}
\label{eq:Birkhoff}
\frac{H(1) + \ldots + H(T)}T \to c,\quad T \to \infty.
\end{equation}
Immediately from~\eqref{eq:Birkhoff}, we have:
\begin{equation}
\label{eq:Birkhoff-shifted}
\frac{H(0) + \ldots + H(T-1)}T \to c,\quad T \to \infty.
\end{equation}
We get~\eqref{eq:conv-to-0} immediately from~\eqref{eq:Birkhoff} and~\eqref{eq:Birkhoff-shifted}. This completes the proof of ~\eqref{eq:limiting-delta}. Similarly,~\eqref{eq:R-LLN} follows from~\eqref{eq:limiting-delta} and~\eqref{eq:conv-to-0}.

\subsection{Proof of Theorem~\ref{thm:Markov}} Define the density of 
$\mathcal N(0, \sigma^2)$: 
\begin{equation}
\label{eq:Gauss}
\varphi(z) = \frac1{\sqrt{2\pi}\sigma}e^{-\frac{z^2}{2\sigma^2}}.
\end{equation}
Then we can express $\varepsilon(t) = H(t) - h - b(H(t-1) - h)$ using~\eqref{eq:main-reg}. Therefore, the transition density of $(G, H)$: condition density of $(G(t), H(t))$ at $(g_1, h_1)$ given $G(t-1) = g_0$ and $H(t-1) = h_0$, is given by
\begin{equation}
\label{eq:transition}
\mathbf{p}(g_0, h_1 - h - b(h_0 - h), g_1)\varphi(h_1 - h - b(h_0 - h)).
\end{equation}

\subsection{Proof of Theorem~\ref{thm:ergodic}}{\it Step 1.}  The function $\mathbf{p}$ from Assumption~\ref{asmp:Markov} is strictly positive for any argument values. The Gaussian density~\eqref{eq:Gauss} is strictly positive too. Thus, the same is true for the function~\eqref{eq:transition}. Therefore, the Markov process $(G, H)$ has the {\it positivity property:} For a subset $A \subseteq \mathbb R^2$ of positive Lebesgue measure, and for any $g_0, h_0 \in \mathbb R^2$, 
\begin{equation}
\label{eq:pve}
\mathbb P((G(1), H(1)) \in A\mid G(0) = g_0,\, H(0) = h_0) > 0.
\end{equation}

{\it Step 2.} Solve the recurrent equation~\eqref{eq:main-reg}: $H(t) - h = b^{t-1}\varepsilon(1) + \ldots + \varepsilon(t) + b^t(H(0) - h)$. This random variable has the same distribution as
$$
H'(t) = \sum\limits_{s=1}^tb^{s-1}\varepsilon(s) + b^t(H(0) - h) + h.
$$
The following series converges in $L^p$:
\begin{equation}
\label{eq:L-p}
H'(\infty) := \sum\limits_{s=1}^{\infty}b^{s-1}\varepsilon(s).
\end{equation}
Indeed, the $p$-norm of $\varepsilon(s)$ is constant (does not depend on $s$), and $b \in (0, 1)$. Therefore, the series~\eqref{eq:L-p} converges in $L^p$. We apply that the space $L^p$ is {\it Banach}, that is, every fundamental sequence converges.
Thus the sequence $H'(t) \to H'(\infty) + h$ as $t \to \infty$ in $L^p$. Therefore, the distribution of $H(t)$ (the same as that of $H'(t)$) converges in law to the distribution of $H'(\infty) + h$. In addition, the $p$th moments of $H(t)$ converges to the $p$th moment of $H'(\infty) + h$ as $t \to \infty$. Since convergence in $\mathcal W_p$ is equivalent to weak convergence plus convergence of $p$th moments, the above results implies convergence in $\mathcal W_p$. Finite $p$th moments of the stationary distribution $\Pi$ are implied by~\eqref{eq:bdd-g-h} and Fatou's lemma. We get the marginal distribution of $\Pi$ for $H$ from the observation that $H$ is an autoregression of order 1 with Gaussian innovations; see any classic text on time series, for example \cite{BrockwellDavis}. 

{\it Step 3.} Finally, let us show uniqueness of the stationary distribution and convergence in total variation. This follows from the two observations. First, as mentioned above, the process $(G, H)$ has the positivity property~\eqref{eq:pve}. Second, we have:
\begin{equation}
\label{eq:bdd-g-h}
\sup_t\mathbb E\left[|G(t)|^p + |H(t)|^p\right] < \infty.
\end{equation}
For $G$, this is in Assumption~\ref{asmp:bdd}; and for $H$, this was shown above, since $H'(t)$ converges in $L^p$ as $t \to \infty$. From~\eqref{eq:bdd-g-h}, we get that the sequence $(G(t), H(t))$ is {\it tight:}
\begin{equation}
\label{eq:tight}
\sup\limits_{t = 0, 1, \ldots}\left[\mathbb P(|G(t)| > u) + \mathbb P(|H(t)| > u)\right] \to 0,\quad u \to \infty.
\end{equation}
In \cite{MT1992}, the property~\eqref{eq:tight} is called {\it boundedness in probability}. This process $(G, H)$ is also irreducible (that is, it does not have two or more disconnected components of the state space $\mathbb R^2$), since the transition density is strictly positive everywhere. The function $\varphi$ from~\eqref{eq:Gauss} is continuous. The function $\mathbf{p}$ from Assumption~\ref{asmp:Markov} is continuous in the first two arguments. Therefore, the Markov process $(G, H)$ is a {\it $T$-chain} in the sense of \cite[p.548]{MT1992}. Applying Corollary (ii) in \cite[p.550]{MT1992}, we conclude that the Markov process $H$ is {\it positive Harris recurrent} in the sense of \cite{MT1992}. Therefore, it has a unique stationary distribution. Applying \cite[Theorem 3.4]{MT1992} to Lebesgue measure $\psi$ (since $(G, H)$ is irreducible with respect to this Lebesgue measure), we get: all compact sets are petite. Apply \cite[Theorem 5.2(iii)]{MT1992}. Use the aperiodicity ($m = 1$ in the notation on the cited theorem), which follows from positivity of the transition density. This completes the proof of Step 3.

\subsection{Proof of Theorem~\ref{thm:CLT}} We rewrite
\begin{align*}
&\frac{Q(1) + \ldots + Q(T) - (c+g)T}{\sqrt{T}} = \\ & \frac{G(1) + \ldots + G(T) - gT}{\sqrt{T}} + \frac{H(T) - H(0)}{\sqrt{T}}.
\end{align*}
Recall Assumption~\ref{asmp:CLT}. By the Slutsky theorem, see for example \cite[Theorem 5.5.17]{Casella}, it suffices to show convergence in distribution (or, equivalently, in probability): 
\begin{equation}
\label{eq:conv-to-0-sqrt}
\frac{H(t) - H(0)}{\sqrt{t}} \to 0,\quad t \to \infty.
\end{equation}
But $H(t) \to H(\infty) \sim \pi_H$ weakly converges to the stationary distribution. Applying the Slutsky theorem again, $H(t)/\sqrt{t} \to 0$ in probability. This proves~\eqref{eq:conv-to-0-sqrt}. 

\subsection{Proof of Theorem~\ref{thm:withdrawal}} (a) We can rewrite 
$$
V_W(t) = V_W(t-1)e^{Q(t) - w^*},\quad t = 1, 2, \ldots
$$
where $w^* = -\ln(1 - w) < g + c$. By induction, $V_W(t) = \exp\left[Q(1) + \ldots + Q(t) - tw^*\right]$. We can represent this exponent as 
$$
\ln V_W(t) = G(1) + \ldots + G(t) + H(t) - H(0) + ct - w^*t.
$$
Using an earlier result~\eqref{eq:conv-to-0}, we get:
$$
\frac1t\ln V_W(t) = \frac1t\sum\limits_{k=1}^tG(k) + \frac1tH(t) - \frac1tH(0) + c - w^* \to g + c - w^* > 0.
$$
Thus $V_W(t) \to \infty$ a.s. as $t \to \infty$. 

\bigskip

(b) Similarly,  we can represent 
$$
\ln V_W(t) = G(1) + \ldots + G(t) + H(t) - H(0) + ct + \ln(1 - W(1)) + \ldots + \ln(1 - W(t)).
$$
Note that $-\ln(1 - w) \ge w$ for $w \in (0, 1)$. Thus 
$$
\ln V_W(t) \le G(1) + \ldots + G(t) + H(t) - H(0) + ct - W(1) - \ldots - W(t).
$$
Dividing by $t$ and letting $t \to \infty$, we have a.s. convergence of the right-hand side:
$$
\frac1t\ln V_W(t) \le \frac1t\sum\limits_{k=1}^tG(k) + \frac{H(t)}t - \frac{H(0)}t + c  \to g + c - w < 0.
$$
Thus $\varlimsup\limits_{t \to \infty} V_W(t) \le 1$, and this withdrawal rate is not sustainable. 

\subsection{Proof of Lemma~\ref{lemma:new-withdraw}} The corresponding wealth process satisfies the recurrent relation:
$$
V(t) = V(t-1)e^{Q(t)}(1 - W(t)) = V(t-1)e^{Q(t) - G(t) - u}.
$$
Applying this recurrent relation multiple times, we get:
\begin{align}
\label{eq:wealth-selected}
\begin{split}
V(t) &= \exp\left[Q(1) + \ldots + Q(T) - G(1) - \ldots G(T) - Tu\right] \\ & = \exp\left[H(T) - H(0) + (c-u)T\right].
\end{split}
\end{align}
Combine~\eqref{eq:wealth-selected} with an earlier result~\eqref{eq:conv-to-0}, we get the following: 

\begin{itemize}
\item If $c > u$, we get: $H(T) + (c-u)T \to +\infty$ as $T \to \infty$, and therefore $V(t) \to +\infty$. The withdrawal process is sustainable.
\item If $c < u$, then $H(T) + (c-u)T \to -\infty$ as $T \to \infty$, therefore $V(T) \to 0$, and the withdrawal process is not sustainable. 
\item Finally, if $c = u$, then the autoregressive process $H(t)$ does not have an almost sure limit, so the same is true for $V(t)$. The withdrawal process is not sustainable. 
\end{itemize}


\begin{thebibliography}{99}

\bibitem{Arnott} \textsc{Robert D. Arnott, Denis B. Chaves, Tzee-man Chow} (2017). King of the Mountain: The Shiller P/E and Macroeconomic Conditions. \textit{The Journal of Portfolio Management} \textbf{44} (1), 55--68.

\bibitem{DividendSurvey} \textsc{H. Kent Baker, Robert A. Weigand} (2015). Corporate Dividend Policy Revisited. \textit{Managerial Finance} \textbf{41} (2), 126--144. 

\bibitem{JPMorgan} \textsc{Alicia Barrett, Peter Rappoport} (2011). Price-Earnings Investing. \textit{JP Morgan Asset Management, Reality in Returns} \textbf{November 2011} (1), 1--12.

\bibitem{BrockwellDavis} \textsc{Peter J. Brockwell, Richard A. Davis} (2016). \textit{Introduction to Time Series and Forecasting.} 3rd edition, Springer.

\bibitem{Bunn} \textsc{Oliver D. Bunn, Robert J. Shiller} (2014). Changing Times, Changing Values: A Historical Analysis of Sectors within the US Stock Market 1872--2013. NBER Working Paper 20370.

\bibitem{Shiller1998} \textsc{John Y. Campbell, Robert J. Shiller} (1998). Valuation Ratios and the Long-Run Stock Market Outlook. \textit{The Journal of Portfolio Management} \textbf{24} (2), 11--26.

\bibitem{Casella} \textsc{George Casella, Roger L. Berger} (2002). \textit{Statistical Inference.} 2nd edition, Cengage. 

\bibitem{FF1993} \textsc{Eugene F. Fama, Kenneth R. French} (1993). Common Risk Factors in the Returns on Stocks and Bonds. \textit{The Journal of Financial Economics} \textbf{33} (1), 3--56.

\bibitem{Farouk} \textsc{Farouk Jivrai, Robert J. Shiller} (2017). The Many Colours of CAPE. Yale ICF Working Paper No. 2018-22.

\bibitem{RW12} \textsc{Burton Malkiel} (2023). \textit{A Random Walk Down Wall Street.} 13th edition, W. W. Norton \& Company.

\bibitem{MT1992} \textsc{Sean P. Meyn, Richard L. Tweedie} (1992). Stability of Markovian Processes I: Criteria for Discrete-Time Chains. \textit{Advances in Applied Probability} \textbf{24}, 542--575. 

\bibitem{MM61} \textsc{Merton H. Miller, Franco Modigliani} (1961). Dividend Policy, Growth, and the Valuation of Shares. \textit{The Journal of Business} \textbf{34} (4), 41--433. 

\bibitem{Acct} \textsc{Jane A. Ou, Stephen H. Penman} (1989). Accounting Measurement, Price-Earnings Ratio, and the Information Content of Security Prices. \textit{Journal of Accounting Research} \textbf{27}, 111--144.

\bibitem{Ural} \textsc{Thomas Philips, Cenk Ural} (2016). Uncloaking Campbell and Shiller’s CAPE: A Comprehensive Guide to Its Construction and Use. \textit{The Journal of Portfolio Management} \textbf{43} (1), 109--125.

\bibitem{Capital} \textsc{Thomas Piketty} (2017). {\it Capital in the Twenty-First Century}. {\it Belknap Press}

\bibitem{PE} \textsc{Pu Shen} (2000). The P/E Ratio and Stock Market Performance. \textit{Federal Reserve Bank of Kansas City Ecomonic Review} \textbf{Q} (4), 23--36.

\bibitem{ShillerVol} \textsc{Robert J. Shiller} (1992). \textit{Market Volatility}. MIT University Press.

\bibitem{ShillerBook} \textsc{Robert J. Shiller} (2015). \textit{Irrational Exuberance.} Third edition, Princeton University Press.

\bibitem{SiegelBook} \textsc{Jeremy G. Siegel} (2022). \textit{Stocks for the Long Run.} Sixth edition, McGraw-Hill.

\bibitem{Stroock} \textsc{Daniel Stroock} (2011). \textit{Probability Theory: An Analytic View}. Second edition. Cambridge University Press.

\end{thebibliography}
\end{document}